\documentclass[]{webofc}
\usepackage[varg]{txfonts}   % Web of Conferences font
\def\msol{\hbox{\kern 0.20em $M_\odot$}}
\def\lsol{\hbox{\kern 0.20em $L_\odot$}}
\def\rsol{\hbox{\kern 0.20em $R_\odot$}}
\def\sr{\hbox{\kern 0.20em sr}}
\def\srmu{\hbox{\kern 0.20em sr$^{-1}$}}
\def\g{\hbox{\kern 0.20em g}}
\def\gmu{\hbox{\kern 0.20em g$^{-1}$}}
\def\kg{\hbox{\kern 0.20em kg}}
\def\pc{\hbox{\kern 0.20em pc}}
\def\mum{\hbox{\kern 0.20em $\mu$m}}
\def\mumd{\hbox{\kern 0.20em $\mu$m$^{-2}$}}
\def\cm{\hbox{\kern 0.20em cm}}
\def\m{\hbox{\kern 0.20em m}}
\def\km{\hbox{\kern 0.20em km}}
\def\nm{\hbox{\kern 0.20em nm}}
\def\s{\hbox{\kern 0.20em s}}
\def\h{\hbox{\kern 0.20em h}}
\def\sec{\hbox{\kern 0.20em sec}}
\def\min{\hbox {\kern 0.20em min}}
\def\smu{\hbox{\kern 0.20em s$^{-1}$}}
\def\smd{\hbox{\kern 0.20em s$^{-2}$}}
\def\an{\hbox{\kern 0.20em an}}
\def\anmu{\hbox{\kern 0.20em an$^{-1}$}}
\def\deg{\hbox{\kern 0.20em $^{\rm o}$}}
\def\yr{\hbox{\kern 0.20em yr}}
\def\yrmu{\hbox{\kern 0.20em yr$^{-1}$}}
\def\Myr{\hbox{\kern 0.20em Myr}}
\def\Mymu{\hbox{\kern 0.20em Myr$^{-1}$}}
\def\K{\hbox{\kern 0.20em K}}
\def\pcmu{\hbox{\kern 0.20em pc$^{-1}$}}
\def\pcmd{\hbox{\kern 0.20em pc$^{-2}$}}
\def\pcmt{\hbox{\kern 0.20em pc$^{-3}$}}
\def\kms{\hbox{\kern 0.20em km\kern 0.20em s$^{-1}$}}
\def\kmpd{\hbox{\kern 0.20em km$^{2}$}}
\def\kpc{\hbox{\kern 0.20em kpc}}
\def\cms{\hbox{\kern 0.20em cm\kern 0.20em s$^{-1}$}}
\def\erg{\hbox{\kern 0.20em erg}}
\def\ergs{\hbox{\kern 0.20em erg}}
\def\cmpd{\hbox{\kern 0.20em cm$^2$}}
\def\cmmd{\hbox{\kern 0.20em cm$^{-2}$}}
\def\cmms{\hbox{\kern 0.20em cm$^{-6}$}}
\def\cmpt{\hbox{\kern 0.20em cm$^3$}}
\def\cmmt{\hbox{\kern 0.20em cm$^{-3}$}}
\def\mpd{\hbox{\kern 0.20em m$^2$}}
\def\mmd{\hbox{\kern 0.20em m$^{-2}$}}
\def\mpt{\hbox{\kern 0.20em m$^3$}}
\def\mmt{\hbox{\kern 0.20em m$^{-3}$}}
\def\mujy{\hbox{\kern 0.20em $\mu$Jy}}
\def\mjy{\hbox{\kern 0.20em mJy}}
\def\Mj{\hbox{\kern 0.20em MJy}}
\def\jy{\hbox{\kern 0.20em Jy}}
\def\ghz{\hbox{\kern 0.20em GHz}}
\def\srmd{\hbox{\kern 0.20em sr$^{-1}$}}

\def \mum{$\mu$m}
\def\G{\hbox{\kern 0.20em G}}

\def\h13cop{\hbox{H$^{13}$CO$^{+}$}}

\def\h2o{\hbox{H$_2$O}}

\begin{document}
\title{The emergence of molecular complexity in star forming regions
  as seen with ASAI}

\author{\firstname{Bertrand} \lastname{Lefloch}\inst{1}\fnsep\thanks{\email{bertrand.lefloch@univ-grenoble-alpes.fr}} \and
  \firstname{Charlotte} \lastname{Vastel}\inst{2,3}
  % \fnsep\thanks{\email{Mail address for second author if necessary}}
  \and
  \firstname{Eleonora} \lastname{Bianchi}\inst{1}
  % \fnsep\thanks{\email{Mail address for last  author if necessary}}
  \and
  \firstname{Rafael} \lastname{Bachiller}\inst{4}
}

\institute{Univ. Grenoble Alpes, CNRS, IPAG, 38000 Grenoble, France
\and
Université de Toulouse, UPS-OMP, IRAP, Toulouse, France
\and
CNRS, IRAP, 9 Av. Colonel Roche, BP 44346, 31028 Toulouse Cedex 4, France
\and
 IGN, Observatorio Astrónomico Nacional, Apartado 1143, 28800 Alcalá
 de Henares, Spain
          }

\abstract{
 The Large Program "Astrochemical Surveys At IRAM" (ASAI) investigates the emergence of molecular complexity 
 along the different stages of the solar-type  star formation process, by carrying out  unbiased  line surveys of 
 a sample of ten template sources in the range 80-272 GHz with the IRAM 30m telescope. We present here an overview of the main results of the Large Program ASAI. }
\maketitle
\section{The motivation for ASAI}
\label{intro}

The chemical evolution of the matter during the long process that brought it
from prestellar cores (PSCs), to protostars and their associated shocks, protoplanetary disks, and ultimately to the bodies of the Solar System 
is one of the key questions in modern Astrophysics and it was  the main motivation for the Large Program "Astrochemical Surveys At IRAM"  (ASAI; \cite{Lefloch2018}) at the IRAM 30m telescope. More precisely, ASAI targeted two specific goals~: a)~to obtain an evolutionary view 
on chemistry, in particular Complex Organic Molecules (COMs) \citep{Herbst2009}: molecular abundances, formation and destruction pathways, b)~to understand the influence of the environmental conditions, in particular the star formation feedback processes. 
Using the EMIR receivers, ASAI  carried out  unbiased line surveys of the molecular line emission of a sample of ten template sources illustrating the different chemical stages of a Sun-like  star during its formation process, from the cold prestellar phase (PSC TMC1 and L1544), to solar-type protostars (First HydroStatic Core B1b, Class 0 IRAS4A and L1157mm, Class 0/I L1527, Class I SVS13A), outflows (L1157-B1, L1448-R2), and last to protoplanetary disks (Class II AB Aur). The observations covered the range 80--272 GHz for all sources but PSCs (70-110 GHz for L1544; 130-170 GHz for TMC1). They were carried out over 6 semesters between 2012 and 2015 (360 hours).

\section{Overall results}

The 3mm window was first investigated in full detail because of its high SNR. The number of the detected species (main isotopologues)  is $\approx 45$, with low variations between the sources of the sample~: 50 in L1544, 51 in B1b, 47 L1527, 48 in IRAS4A, 35 in SVS13A.  This number  appears rather independent of the source evolutionary stage and luminosity, meaning that chemical complexity is present at all stages of protostellar evolution. Similar numbers of molecular species are actually found in the massive SFRs Orion (43) and SgrB2 (56). The main difference with the more luminous and massive sources resides in the spectral line density, which is lower by one order of magnitude in  solar-type sources ($\sim 5$-12 GHz$^{-1}$). No other COMs than those previously known  were detected with ASAI.

\begin{figure}[h]
\centering
\includegraphics[width=0.9\columnwidth,clip]{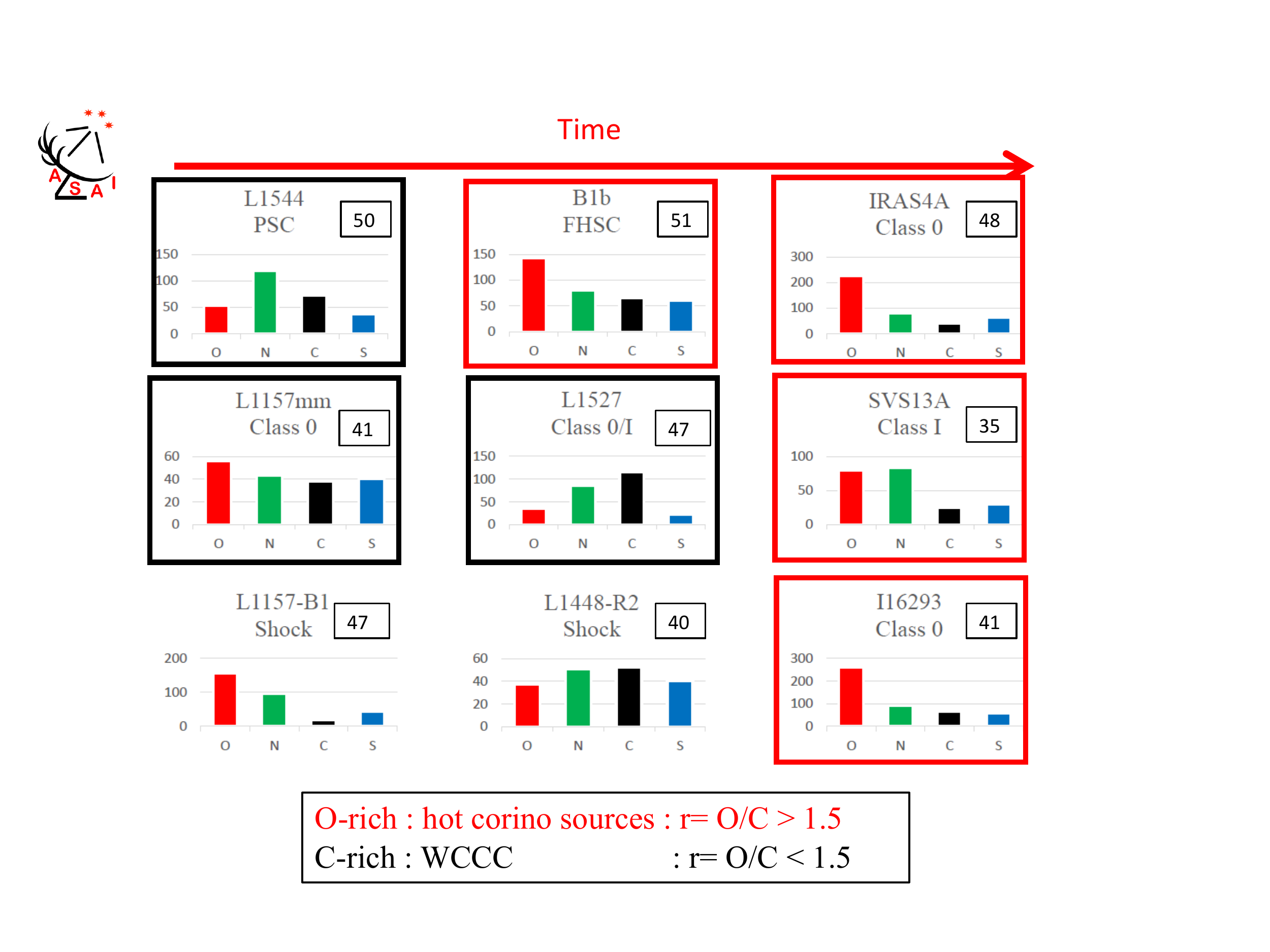}
\caption{Distribution of the composition in chemical families based on the number of molecular lines in the 3mm band: Oxygen, Nitrogen, Carbon, Sulfur. Similar results are obtained when considering the number of molecular species. The ratio  r of O- to C-bearing species allows to determine if protostars are hot corino (r > 1.5) or WCCC sources (r < 1.5). The evolutionary status and the number of species (main isotopologues) are given for each source. Adapted from \cite{Lefloch2018}. }
\label{fig-3}       % Give a unique label
\end{figure}

Very significant differences are observed between the spectra of the sample sources, reflecting the evolutionary stage and the physical conditions
%, as illustrated in Fig.~1. 
This chemical differentiation  affects not only deuteration or S-bearing species but also COMs. To determine how much the molecular content differs between sources, an inventory of the molecular species by "elemental families" was obtained: 
%(hydro)carbon (CxHy), oxygen (CxHyOz), nitrogen (CxHyOzNt), %sulfur (CxHyOzNtSu), silicon (Si-), phosphorus (P-). 
C, O, N, S, Si, P
It allowed to classify the source sample  into two main chemical families,  depending on the ratio of O-bearing to C-bearing species: first, hot corinos (prototype~: IRAS16293 -- 2422) have a rich content in oxygen with respect to carbon; second, sources 
with a rich content in  hydrocarbon  with respect to oxygen; it is the signature of the Warm Carbon Chain Chemistry (WCCC) sources (prototype~: L1527). The ratio r of O-bearing to hydrocarbon species allows to discriminate between hot corinos 
(r= 2-4) and  WCCC sources (r= 1-1.4) in the sample. The question remains whether another chemical class  (N- or /S-) could arise depending on the environmental/initial conditions of star formation. A study similar to ASAI on a broader sample of sources and SFRs is needed to address this question.

\section{Cold but not poor in organics}
%\label{sec:PSC}
%

\begin{figure}[h]
\centering
\sidecaption
\includegraphics[width=0.49\textwidth,clip]{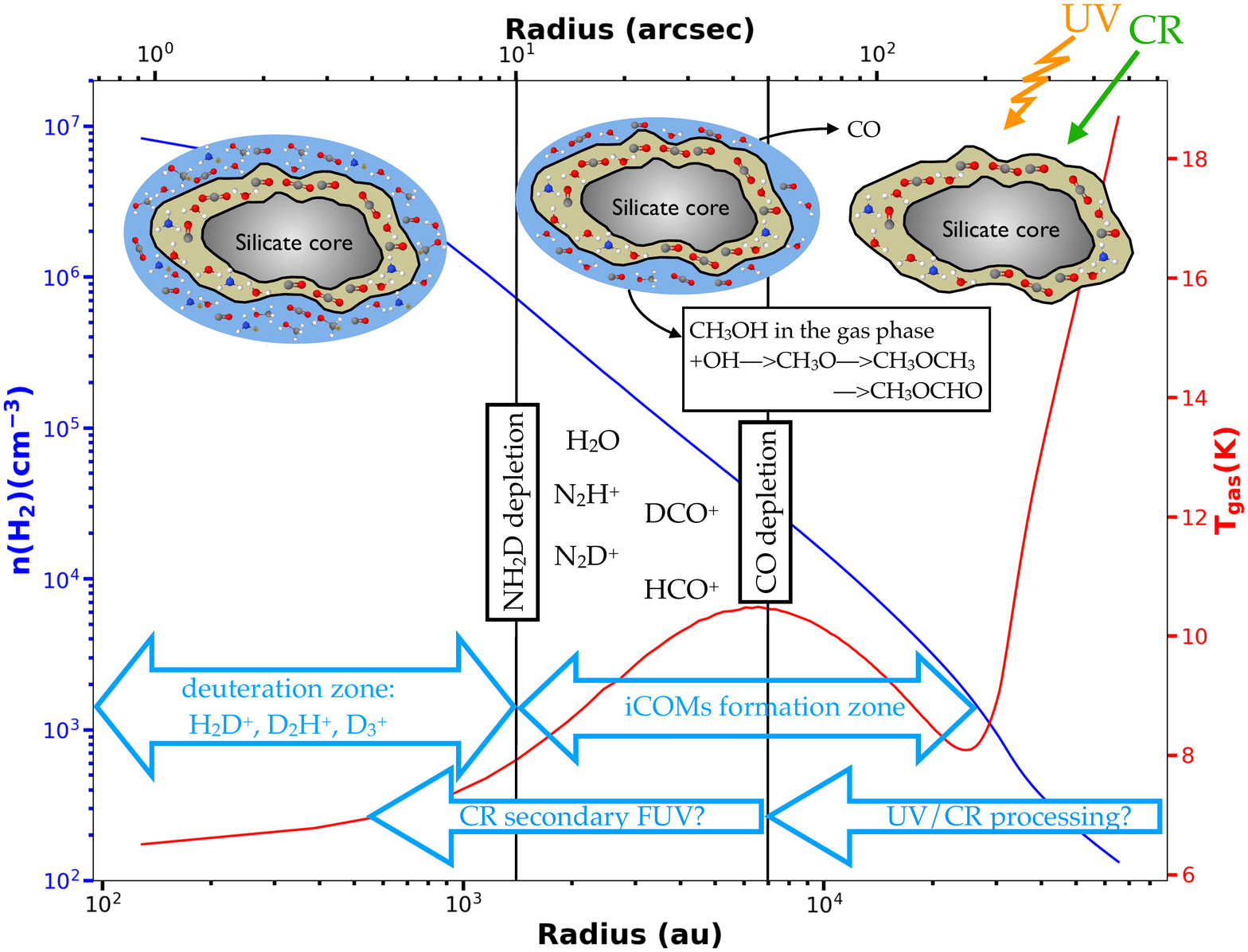}
\caption{Sketch  of the physical-chemical structure of the prototypical prestellar core L1544 (from the review by \cite{Ceccarelli2022} with the permission of the authors). The dust grains are illustrated, as silicate cores, covered with the grain surface  chemical composition (yellow) coated with the icy grain mantle (in blue; thicker in the cold dense interiors of the core).  The temperature bump at $\sim$ 7000 au corresponds to the so-called photo-desorption layer, a chemical zone between the FUV starlight pervading interstellar space and the FUV produced by cosmic-rays.
}
\label{l1544}       % Give a unique label
\end{figure}

L1544 is one of the densest and most dynamically evolved starless cores known.
%, towards which was detected H$_2$D$^+$ \citep{Caselli2003} and H$_{2}$O %\citep{Caselli2012b} for the first time in the cold ISM. 
Its general physical-chemical structure is shown in the scheme of 
Fig.~\ref{l1544}. On this Figure, the two vertical black lines show the radius below which CO and NH$_2$D start to freeze-out onto the grain mantles, disappearing from the gas-phase. \\

The L1544 survey ranges from 70 to 110 GHz and led to the detection of a very rich chemistry. For instance,  more than twenty sulphur bearing species have been detected with more than 50 transitions \citep{Vastel2018a,Cernicharo2018} such as CS, CCS, C$_3$S, SO, SO$_2$, H$_2$CS, OCS, HSCN, NS, HCS$^+$, NS$^+$, H$_2$S and CH$_3$SH. From the chemical modelling they conclude that these species present a strong radial variation for the abundances, which mostly are emitted in the external layer where non thermal desorption of other species has previously been observed. They 
concluded that a strong sulphur depletion is necessary to fully reproduce the observations. Within ASAI many nitrogen bearing species have also been detected such as the hyperfine structure of the cyanomethyl radical CH$_2$CN \citep{Vastel2015}, HNC$_3$, HCCNC and HC$_3$N \citep{Vastel2018b,Hily-Blant2018}, HC$_3$NH$^+$ and HCNH$^+$ \citep{Quenard2017}, CNCN, NCCNH$^+$, C$_3$N, CH$_3$CN, C$_2$H$_3$CN, and H$_2$CN \citep[][]{Vastel2019}. Molecular ions have also been studied in ASAI such as HOCO$^+$ \citep{Vastel2016}, leading to an indirect estimate of CO$_2$ in the gas phase in the [$2 \times 10^{-7}-2 \times 10^{-6}$] range. And COMs have also been detected \citep{Vastel2014,Jimenez2016} such as methanol (CH$_3$OH), acetaldehyde (CH$_3$CHO) and dimethyl ether (CH$_3$OCH$_3$) with integrated intensities of just a few mK~km s$^{-1}$. From these ASAI detections, the above authors built a detailed chemical network to reproduce the molecular abundances and found that most emission occur in the external envelope of the prestellar core \citep{Vastel2014,Vasyunin2017}, while the simple deuterated species trace the internal structure where CO is depleted \citep[e.g.][]{Sipila2018}. The non-LTE analysis of the methanol lines \citep{Vastel2014} concluded that the bulk of the emission originates in the external ($\sim$ 8000 au from the center), relatively dense ($\sim$ 2 $\times$ 10$^4$ cm$^{-3}$) and cold ($\sim$ 10 K) layers of L1544, what is defined as molecular and COMs zone in Fig.~\ref{l1544}, through non-thermal processes. Over the years, a total of about 140 molecules have been detected towards L1544, one fourth being COMs.

The molecular complexity found in L1544 was surprising at first,  since the low temperatures and high densities did not favour neither an active surface chemistry nor the presence of COMs in the gas phase, usually highly bounded to the dust grain surfaces. COMs were indeed mainly found in warm environments, such as hot cores and hot corinos (see Sect.~4). 
%At the low temperatures found in prestellar cores, only the light atoms such as hydrogen and deuterium atoms can sweep across dust grain surfaces, thus allowing them to easily react with other species. 
Many detailed astrochemical models have been developed over the past few years to predict the COMs abundances in a layer at about 8000 au found in our ASAI observations. We refer to the Protostars and Planets VII chapter from \cite{Ceccarelli2022} for a review on the models that aim to reproduce the molecular complexity in the cold regions of the ISM.

\section{Solar-type protostars}
\label{sec:protostars}

\textbf{A Class I hot corino: }

\begin{figure}[h]
\centering
\sidecaption
\includegraphics[width=0.73\textwidth,clip]{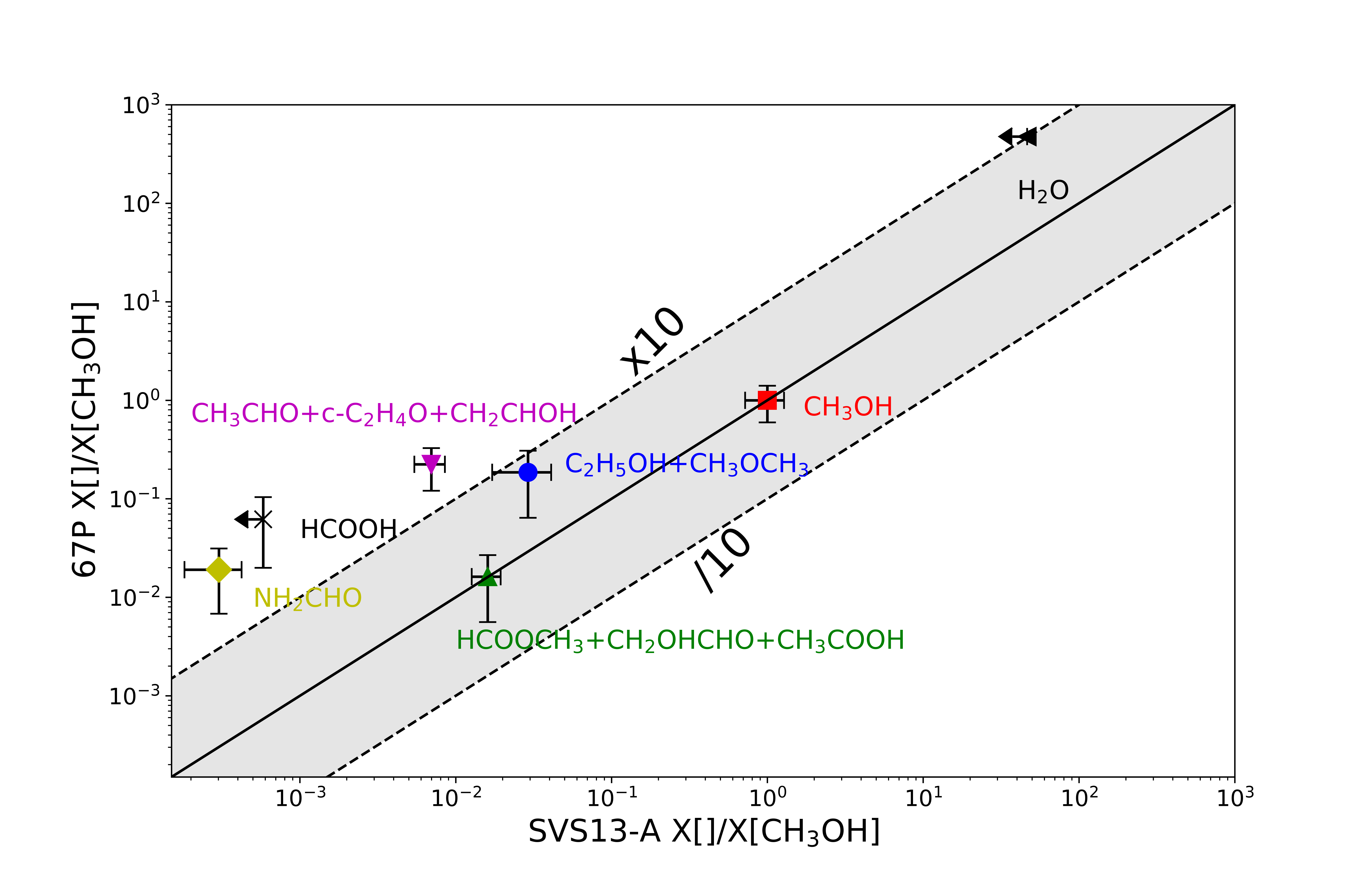}
\hspace{-1cm}
\caption{Abundance ratios, normalized to methanol, of COMs detected in SVS13A and in the 67P comet \citep[adapted from][]{Bianchi2019}. Abundances of isomers are not distinguished since in the 67P the abundances are measured using a mass spectrometer.}
\label{fig3}       % Give a unique label
\end{figure}

The ASAI survey revealed a rich hot corino chemistry in the Class I protostar SVS13A, thanks to the detection of more than 100 emission lines from interstellar COMs such as acetaldehyde (CH$_3$CHO), methyl formate (HCOOCH$_3$), dimethyl ether (CH$_3$OCH$_3$), ethanol (C$_2$H$_5$OH) and, formamide (NH$_2$CHO), in addition to other less complex organics like methanol (CH$_3$OH), methyl cyanide (CH$_3$CN), cyanoacethylene (HC$_3$N), formaldehyde (H$_2$CO) and, thioformaldehyde (H$_2$CS) \citep{Bianchi2017, Bianchi2019, Bianchi2019ACS, Bianchi2022}. 
The COMs abundance ratios are in good agreement with those measured in younger Class 0 sources such as IRAS 4A, suggesting that Class I hot corinos are as chemically rich as younger Class 0 ones  \citep{Bianchi2019}. These results provided a first evidence of chemical inheritance between the Class 0 and the Class I phase.
Moreover, the comparison with the measurements in the 67P comet, obtained thanks to the ROSETTA mission, showed that some COMs (ethanol, methyl formate and acetaldehyde) are in agreement within a factor of 10 (see Fig. \ref{fig3}).
This work suggested that these COMs in comets could be at least partially inherited from the protostellar stage \citep{Bianchi2019}.

\begin{figure}[h]
\centering
\sidecaption
\includegraphics[width=0.73\textwidth,clip]{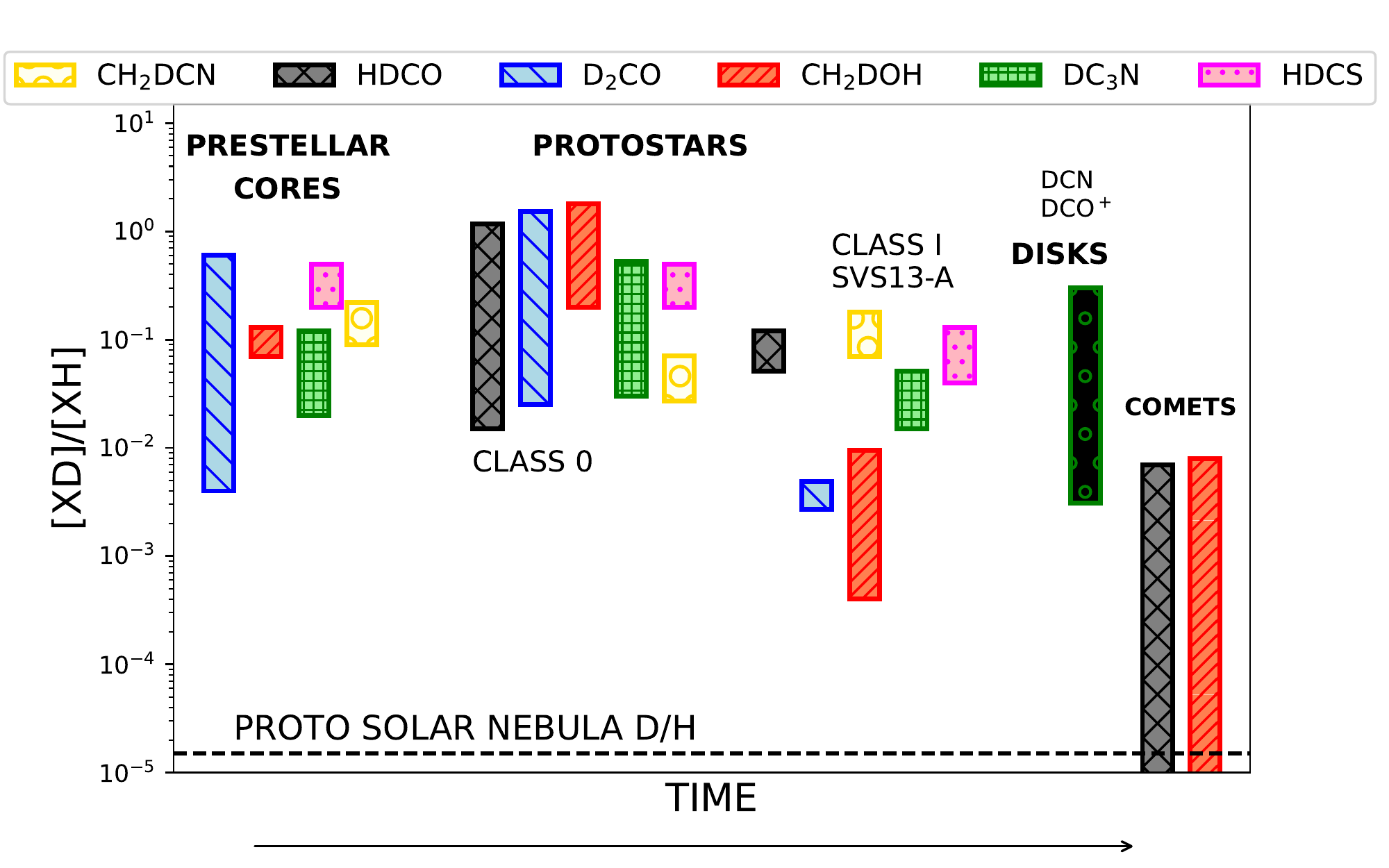}
\caption{[XD]/[XH] ratio measured in organic matter in different astronomical sources: prestellar cores, Class 0 protostars, the Class I SVS13A, and protoplanetary disks \citep[adapted from][]{Bianchi2019ACS}. The measurements towards SVS13A fill the gap between Class 0 protostars and protoplanetary disks.}
\label{fig4}       % Give a unique label
\end{figure}

\textbf{Deuterium fractionation: } 
We have measured the deuterium fractionation of several molecular species in SVS13A and compared it to those obtained in prestellar cores, Class 0 protostars, protoplanetary disks and comets (see Fig. \ref{fig4}). The  comparison suggested again inheritance between the evolutionary stages, with the exception of methanol deuteration which seems to show a significant decrease in the Class I phase.
Moreover, the different degree of deuteration measured between the different species gave us information about their different formation pathways. For example, we could speculate that CH$_3$CN formation takes place probably during the prestellar stage and in gas-phase \citep{Bianchi2022}. The main result of this work was to complement the deuteration measurement in the Class I phase building a bridge between the observation of younger prestellar cores and Class 0 protostars and those in Solar System objects.

\textbf{Formamide across the evolutionary stages: }
A relevant transversal project was focused on formamide (NH$_2$CHO), one key prebiotic molecular species. In particular, formamide was detected in four out of the 10 ASAI sources (IRAS 4A, SVS13A, Cep E, and OMC-2), and in the Class 0 IRAS 16293--2422. This systematic study increased the number of known low and intermediate-mass formamide-emitting protostars to five and it uncovered an almost linear correlation between HNCO and
NH$_2$CHO abundances, holding for several orders of magnitude. These results contributed to shed light on the formamide formation pathways. Indeed, while it was first suggested that HNCO and
NH$_2$CHO were chemically linked, further experimental and theoretical studies have shown that the correlation is probably physical rather than chemical  because the two species are formed through similar physical processes across different environments \cite[][and references therein.]{Lopez-Sepulcre2019}.

\section{Shocks}
\label{sec:shocks}
The prototypical outflow shock region L1157-B1 
%and the driving protostar  L1157-mm 
provides a remarkable example of the chemical richness and differentiation induced by shocks. 
The excellent sensitivity of the survey led to identify a large number  (10--100) of lines from many COMs, some detected for the first time in a shock \cite{Lefloch2017})~: 
CH$_3$OH, H$_2$CCO, C$_2$H$_5$OH, {\bf CH$_3$CHO}, HCOOCH$_3$, {\bf HCOCH$_2$OH}, {\bf CH$_3$OCH$_3$}, t-HCOOH, CH$_3$CN, {\bf C$_2$H$_3$CN}, {\bf HC$_5$N}, {\bf NH$_2$CHO}, {\bf CH$_3$SH} (species in boldface are first detections in a shock). 
The chemical richness discovered in L1157-B1  compares very well to that of protostellar envelopes at the scale of a  few 1000 au probed by the beam of the IRAM 30m.  COM abundances measured in L1157-B1 are usually  a factor 2-10 higher  than those measured in hot corinos. Abundances of O-bearing COMs are typically $\sim 10^{-8}$, i.e. $2\%$-$5\%$ of CH$_3$OH. 

As mentioned in Sect.~4, linear correlations are sometimes observed between molecular species abundances, which has been proposed to trace a common chemical origin between pair members. The study of L1157-B1 has revealed such correlations for various pairs of COMs and organic species \citep{Mendoza2014,Lefloch2017}. The case of glycolaldehyde HCOCH$_2$OH and ethanol C$_2$H$_5$OH, measured in low-mass protostars and shocks, led to explore a new formation pathway of the former species, poorly constrained until then.  
%A  first possibility is that  glycolaldehyde forms from radical %recombination at the surface of dust grains \cite{Garrod2008}, and %is subsequently released in the gas phase through grain sputtering. 
Based on the ASAI results, \cite{Skouteris2018} proposed a scheme in which glycolaldehyde forms in the gas phase from ethanol, once the latter is released in the gas phase either through sputtering (shock) or sublimation (hot core) of grain mantles.  Interestingly, this new scheme successfully accounts for the observed correlation HCOCH$_2$OH/C$_2$H$_5$OH towards hot corinos and L1157-B1  \citep{Lefloch2017}.

Molecular species with a peptide link -NH-C(=O)-  NH$_2$CHO and HNCO were detected in L1157-B1 with  abundances  as high as those measured in  the most abundant Galactic sources \cite{Mendoza2014}. The importance of these species in the synthesis of amino acids has  stimulated a lot of observational and theoretical work on their formation route. 
The definite answer however came from NOEMA observations 
which unambiguously showed that the NH$_2$CHO emission originates from the post-shock gas \cite{Codella2017} and not from the sputtering region in the shock where the grain mantle material is released into the gas phase.

\section{Conclusion}
ASAI was a very successful project which involved about 30 scientists from 7 laboratories in France, Spain, Italy, UK and Japan. It is worth noticing that it also stimulated and contributed to  theoretical and experimental works in Astrochemistry. 
The ASAI Legacy has provided a vast observational data set which was the basis for the successful Large Program SOLIS  at the IRAM-NOEMA interferometer. Some of the results presented here show that the synergy between the two Large Programs is extremely powerful for the characterization of Solar-type protostars as it allows at the same time to resolve the molecular emission, thanks to the high-angular resolution of SOLIS and to accurately derive the physical gas parameters thanks to the detection of a large number of molecular transitions provided by ASAI.

%
% BibTeX or Biber users please use (the style is already called in the class, ensure that the "woc.bst" style is in your local directory)
% \bibliography{name or your bibliography database}

\end{document}